\RequirePackage{fix-cm}
\documentclass[twocolumn,epjc3]{svjour3}
\smartqed  
\RequirePackage{graphicx}
\journalname{Eur. Phys. J. C}
\begin{document}

\title{Vaidya-Tikekar Type Superdense Star Admitting Conformal Motion in Presence of Quintessence Field }


\author{Piyali Bhar\thanksref{e1,addr1}}

\thankstext{e1}{e-mail:piyalibhar90@gmail.com}


\institute{Department of
Mathematics, Jadavpur University, Kolkata 700 032, West Bengal,
India \label{addr1}}

\date{Received: date / Accepted: date}

\maketitle

\begin{abstract}
To explain the reason  of accelerated expansion of our universe dark energy is a suitable candidate.
Motivated by this concept in the present paper we have obtained a new model of an anisotropic superdense star which admits conformal motions in presence of quintessence field which is characterized by a parameter $\omega_q $ with $-1<\omega_q<-\frac{1}{3}$.The model has been developed by choosing Vaidya-Titekar ansatz [P C Vaidya and R Tikekar (1982){\it J. Astrophys
.Astron.}~{\bf 3}~325].Our model satisfy all the physical requirements.We have analyze our result analytically as well as
with the help of graphical representation.
\keywords{Quintessence field  \and Vaidya-Tikekar relativistic star \and conformal motion}
\end{abstract}

\section{Introduction}
The study of dark matter and dark energy has become a topic of considerable interest in present decades.The study is not only important from theoretical point of view but also from physical point of view.The reason is  many observational evidences suggest that the expansion of our universe is accelerating.Dark energy is the most acceptable hypothesis to explain this.Work done based on the Cosmic microwave background (CMB) estimated that our universe made up of 68.3\% dark energy, 26.8\% dark matter and 4.9\% ordinary matter.Dark matter cannot be seen by telescopes but one can infer its evidence from gravitational effects on visible matter and gravitational lensing of background radiation.On the other hand the evidence of dark energy may inferred from measures of large scale wave patterns of mass density.One notable features of the dark energy is that it has a strong negative pressure i.e,the ratio of pressure to density,which is termed as the equation of state parameter $(\omega)$ is negative.The dark energy equation of state is given by $p=\omega \rho$ with $\omega<-\frac{1}{3}$.The dark energy star model has been studied by several authors \cite{Chan1,Chan2,Ghezzi,lobo1,lobo2,saibal,jadav}.If we choose $\omega=-1$ we will get the model of gravastar \cite{mazur01,mazur04,far12a,far12b,far14,usmani11},which is also a dark energy star. $\omega<-1$ is denoted as phantom energy and it violates the null energy condition.Several authors have used phantom equation of state to describe wormhole model \cite{mk1,mk2,lb1,lb2}.Motivated by these previous work we have chosen quintessence dark energy to develop our present model.Where the quintessence field is characterized by a parameter $\omega_q$ with $-1<\omega_q<-\frac{1}{3}$.We have assumed that the underlying fluid is a mixture of ordinary matter and still an unknown form of matter i.e of dark energy type which is repulsive in nature.These two fluids are non-interacting and we have considered the combine effect of these two fluids in our model.Let us assume that the pressure distribution inside the fluid sphere is not isotropic in nature,it can be decomposed into two parts :radial pressure $p_r$ and transverse pressure $p_t$.Where $p_t$ is in the perpendicular direction to $p_r$ and $\Delta=p_t-p_r$ is defined as anisotropic factor.The reason of choosing anisotropic pressure is inspired by the fact that at the core of the superdense star where the density $\sim 10^{15}$ gm/cc. the matter distribution shows anisotropy.\\

In 1982 Vaidya and Titekar \cite{vaidya} proposed a static spherically symmetric model of a superdense star based on an exact solution of Einstein's equations.The physical 3-space $\left\{t = constant\right\}$ of the star is spheroidal,the density of the star is $\sim 2\times 10^{4} gm/cc$, and mass is about four times the solar mass.Several works have been done by using Vaidya-Titekar ansatz. Gupta and Kumar have studied  charged Vaidya-Titekar star in\cite{gupta2}.In this paper the authors have considered a particular form of electric field intensity that has a positive gradient. The said particular form of electric field intensity was used by Sharma \emph{et al.}\cite{sharma}. Komathiraj and Maharaj \cite{maharaj} have also assumed the same expression to model a new type of Vaidya-Titekar type star. Some new closed form solution of  Vaidya-Titekar type star were obtained by Gupta \emph{et al}\cite{gupta3}. Bijalwan and Gupta \cite{gupta4} have taken a more general form of  electric intensity to obtain a new solution of Vaidya-Titekar type stars with charge analogue. In this paper authors have matched their interior interior solution to the exterior R-N metric and they have analyzed their result numerically by assuming suitable values of the chosen parameter.Some other works on Vaidya-Titekar stars are done in \cite{patel,titekar,maharaj1}\\

~~~~~~~~~~~~~In recent past many researchers have worked on conformal motion. Anisotropic stars admitting conformal motion has been studied by Rahaman \emph{et al} \cite{farook10}.Charged gravastar admitting conformal motion has been studied by Usmani {\emph et al} \cite{usmani11}.Relativistic stars admitting conformal motion has been analyzed in\cite{rahaman10}.Isotropic and anisotropic charged spheres admitting a one parameter group of conformal motions was analyzed in \cite{hleon85,leon85,leon85a}.Charged fluid sphere with linear equation of state admitting conformal motion has been studied in \cite{aloma10}.In this paper the authors have also discussed about the dynamical stability analysis of the system.Ray \emph{et al}\cite{ray04,ray07} have given an electromagnetic mass model admitting conformal killing vector.By assuming the existence of a one parameter group of conformal motion Mak \& Harko \cite{mak04} have described an charged strange quark star model. The above author have also discussed conformally symmetric vacuum solutions of the gravitational field equations in the brane-world models \cite{hm05}.In a very recent work Rahaman \emph{et al} \cite{farook14} have described conformal motion in higher dimensional spacetime\\

~~~~To search the natural relationship between geometry and matter through the Einstein's field equations,we generally use inheritance symmetry.The well known inheritance symmetry is the symmetry under conformal killing vectors(CKV) i.e,
\begin{equation}
L_\xi g_{ik}=\psi g_{ik}
\end{equation}
where $L$ is the Lie derivative of the metric tensor which describes the interior gravitational field of a compact star with respect to the vector field $\xi$ and $\psi$ is the conformal factor.It is supposed that the vector $\xi$ generates the conformal symmetry and the metric $g$ is conformally mapped onto itself along $\xi$.Neither $\xi$ nor $\psi$ need to be static even through one consider a static metric.\cite{Harko07,Harko08}.If $\psi=0$ then $(1)$ gives the killing vector,for $\psi=$ constant it gives homothetic vector and if $\psi=\psi(\textbf{x},t)$ then it yields conformal vectors.Moreover note that if $\psi=0$ the underlying spacetime is asymptotically flat which further implies that the Weyl tensor will also vanish.So CKV provides a deeper insight in the spacetime geometry. \\

~~~~~~~~The plan of our paper as follows:In section $2$ we have discussed about interior solution and Einstein field equation.Conformal killing vector and solution of the system have been given in section $3$ and $4$ respectively.Some physical properties of the model is given in sec $5-11$ and we have discussed about some concluding remarks in sec $12$.

\section{Interior Solutions and Einstein field Equation}
To describe a static spherically symmetry spacetime let us consider the line element in the standard form as,
\begin{equation}
ds^{2}=-e^{\nu(r)}dt^{2}+e^{\lambda(r)}dr^{2}+r^{2}(d\theta^{2}+\sin^{2}\theta d\phi^{2})
\end{equation}
Where $\lambda$ and $\nu$ are function of the radial parameter 'r' only.\\
~~~~~~~Now let us assume that our model contains a quintessence like field along with anisotropic pressure.The Einstein Equations can be written as,
\begin{equation}
G_{\mu \nu}=8\pi G (T_{\mu \nu}+\tau_{\mu \nu})
\end{equation}
Where $\tau_{\mu \nu}$ is the energy momentum tensor of the quintes-sence  like field which is characterized by a parameter $\omega_q$ with $-1<\omega_q<-\frac{1}{3}$.Now Kiselev \cite {Kiselev} has shown that the component of this tensor need to satisfy the conditions of additivity and linearity.Considering the different signature used in line elements,the components can be stated as follows:
\begin{equation}
\tau_t^{t}=\tau_r^{r}=-\rho_q
\end{equation}
\begin{equation}
\tau_{\theta}^{\theta}=\tau_{\phi}^{\phi}=\frac{1}{2}(3\omega_q+1)\rho_q
\end{equation}
and the corresponding energy-momentum tensor can be written as,
\begin{equation}
T_{\nu}^{\mu}=(\rho+p_r)u^{\mu}u_{\nu}-p_t g_{\nu}^{\mu}+(p_r-p_t)\eta^{\mu}\eta_{\nu}
\end{equation}
with $u^{i}u_{j} =-\eta^{i}\eta_j = 1 $ and $u^{i}\eta_j= 0$. Here the vector $u_i$ is the fluid 4-velocity and $\eta^{i}$ is the spacelike vector which is orthogonal to $ u^{i}$, $\rho$ is the energy density, $p_r$ and $p_t$ are respectively the radial and the transversal pressure of the fluid.\\
The Einstein field equation assuming $G=1=c$ can be written as
\begin{equation}
e^{-\lambda}\left[\frac{\lambda'}{r}-\frac{1}{r^{2}} \right]+\frac{1}{r^{2}}=8\pi(\rho+\rho_q)
\end{equation}
\begin{equation}
e^{-\lambda}\left[\frac{1}{r^{2}}+\frac{\nu'}{r} \right]-\frac{1}{r^{2}}=8\pi(p_r-\rho_q)
\end{equation}

\[\frac{1}{2}e^{-\lambda}\left[ \frac{1}{2}\nu'^{2}+\nu''-\frac{1}{2}\lambda'\nu'+\frac{1}{r}(\nu'-\lambda')\right]\]
\begin{equation}
~~~~~~~~~~~~~~~~~~~~~~~~~~~~~~~~~=8\pi\left(p_t+\frac{3\omega_q+1}{2}\rho_q
\right)
\end{equation}
\section{conformal killing equation}

The conformal killing equation $(1)$ becomes,
\begin{equation}
L_\xi g_{ik}=\xi_{i;k}+\xi_{k;i}=\psi g_{ik}
\end{equation}

Now using the conformal killing equation to the line element$(2)$ we get the following equations,
\begin{equation}
\xi^{1}\nu'=\psi
\end{equation}
\begin{equation}
\xi^{4}=C_1
\end{equation}
\begin{equation}
\xi^{1}=\frac{\psi r}{2}
\end{equation}
\begin{equation}
\xi^{1}\lambda'+2\xi^{1},_1=\psi
\end{equation}
~~~~~Where $C_1$ is a constant.\

The above four equations consequently gives,
\begin{equation}
e^{\nu}=C_2^{2}r^{2}
\end{equation}
\begin{equation}
e^{\lambda}=\left(\frac{C_3}{\psi}\right)^{2}
\end{equation}
\begin{equation}
\xi^{i}=C_1\delta_{4}^{i}+\left( \frac{\psi r}{2}\right)\delta_1^{i}
\end{equation}
Where $C_2$ and $C_3$ are constants of integrations.\\
Now using equations $(15)-(17)$ to the Einstein field equations$(7)-(9)$ one can obtain
\begin{equation}
\frac{1}{r^{2}}\left[1-\frac{\psi^{2}}{C_3^{2}}\right]-\frac{2\psi \psi'}{r C_3^{2}}=8\pi (\rho+\rho_q)
\end{equation}
\begin{equation}
\frac{1}{r^{2}}\left[\frac{3\psi^{2}}{r^{2}C_3^{2}}-1\right]=8\pi(p_r-\rho_q)
\end{equation}
\begin{equation}
\frac{\psi^{2}}{C_3^{2}r^{2}}+\frac{2\psi \psi'}{r C_3^{2}}=8\pi\left( p_t+\frac{3\omega_q+1}{2}\rho_q\right)
\end{equation}

\section{solution}
To solve the equations (18)-(20) we consider Vaidya-Titekar ansatz\cite{vaidya}
\begin{equation}
e^{\lambda}=\frac{1-K\left(\frac{r^{2}}{R^{2}}\right)}{1-\frac{r^{2}}{R^{2}}}
\end{equation}
It may be noted that the physical $3-$ space $\left\{t = constant\right\}$ of Vaidya-Titekar type star is spheroidal and the geometry of the 3-spheroid is governed by the parameters R and K.Where the parameter $ K <1$.For $K=0$,the hypersurfaces $\left\{t=constant\right\}$ becomes spherical and it gives Schwarzschild interior solution and for  $K=1$ the hypersurfaces $\left\{t=constant\right\}$ becomes flat. The metric function $e^{\lambda}$ is regular at center  and well behaved for $r<R$.

From equation (16) and (21) we get,
\begin{equation}
\psi^{2}=C_3^{2}\frac{R^{2}-r^{2}}{R^{2}-Kr^{2}}
\end{equation}
Using the value of $\psi$ given in equation (22) we can write equations (18)-(20) as follows:
\begin{equation}
(1-K)\frac{3R^{2}-Kr^{2}}{(R^{2}-Kr^{2})^{2}}=8\pi(\rho+\rho_q)
\end{equation}
\begin{equation}
\frac{1}{r^{2}}\left[\frac{2R^{2}-(3-K)r^{2}}{R^{2}-Kr^{2}}\right]=8\pi(p_r-\rho_q)
\end{equation}
\begin{equation}
\frac{1}{r^{2}}\frac{R^{2}-r^{2}}{R^{2}-Kr^{2}}-\frac{2(1-K)R^{2}}{(R^{2}-Kr^{2})^{2}}=8\pi\left( p_t+\frac{3\omega_q+1}{2}\rho_q\right)
\end{equation}
One can note from equation (23)-(25) that we have three equations with four unknowns namely $\rho$,~$p_r$,~$p_t$,~$\rho_q$.\\

To solve the above three equations[(23)-(25)] let us assume that the radial pressure $p_r$ is proportional to matter density $\rho$ i.e,
\begin{equation}
p_r=m\rho,~~~~~~~~~~~0<m<1
\end{equation}
Where $m$ is the equation of state parameter.\\
Solving equation $(23)-(25)$ with help of equation $(26)$ one can obtain
\[\rho=\frac{1}{8\pi (1+m)}\left[(1-K)\frac{3R^{2}-Kr^{2}}{(R^{2}-Kr^{2})^{2}}+\frac{2R^{2}-(3-K)r^{2}}{r^{2}(R^{2}-Kr^{2})}\right]\]
\begin{equation}\end{equation}
\[p_r=\frac{m}{8\pi (1+m)}\left[(1-K)\frac{3R^{2}-Kr^{2}}{(R^{2}-Kr^{2})^{2}}+\frac{2R^{2}-(3-K)r^{2}}{r^{2}(R^{2}-Kr^{2})}\right]\]
\begin{equation}\end{equation}

\[\rho_q=\frac{1-K}{8\pi}\frac{3R^{2}-Kr^{2}}{(R^{2}-Kr^{2})^{2}}-\]
\[~~~~~~\frac{1}{8\pi (1+m)}\left[(1-K)\frac{3R^{2}-Kr^{2}}{(R^{2}-Kr^{2})^{2}}+\frac{2R^{2}-(3-K)r^{2}}{r^{2}(R^{2}-Kr^{2})}\right]\]
\begin{equation}\end{equation}

\pagebreak

\[p_t=\frac{1}{8\pi}\left[\frac{R^{2}-r^{2}}{r^{2}(R^{2}-Kr^{2})}+\frac{2(K-1)R^{2}}{(R^{2}-Kr^{2})^{2}}\right]-
\frac{(3\omega_q+1)}{2}\times\]

\[\frac{(1-K)}{8\pi}\frac{3R^{2}-Kr^{2}}{(R^{2}-Kr^{2})^{2}}-\frac{3\omega_q+1}{16\pi(1+m)}
\frac{(1-K)(3R^{2}-Kr^{2})}{(R^{2}-Kr^{2})^{2}}
\]
\begin{equation}
~~~~~~~~~~~~+\frac{3\omega_q+1}{16\pi(1+m)}\frac{2R^{2}-(3-K)r^{2}}{r^{2}(R^{2}-Kr^{2})}~~~~~
\end{equation}
We denote
\begin{equation}
\rho_{eff}=\frac{1}{8\pi}\left[(1-K)\frac{3R^{2}-Kr^{2}}{(R^{2}-Kr^{2})^{2}}\right]
\end{equation}

\begin{equation}
p_{r~eff}=\frac{1}{8\pi}\frac{1}{r^{2}}\left[\frac{2R^{2}-(3-K)r^{2}}{R^{2}-Kr^{2}}\right]
\end{equation}
\begin{equation}
p_{t~eff}=\frac{1}{8\pi}\left[\frac{1}{r^{2}}\frac{R^{2}-r^{2}}{R^{2}-Kr^{2}}-\frac{2(1-K)R^{2}}{(R^{2}-Kr^{2})^{2}}\right]
\end{equation}

\begin{figure}[htbp]
    \centering
        \includegraphics[scale=.3]{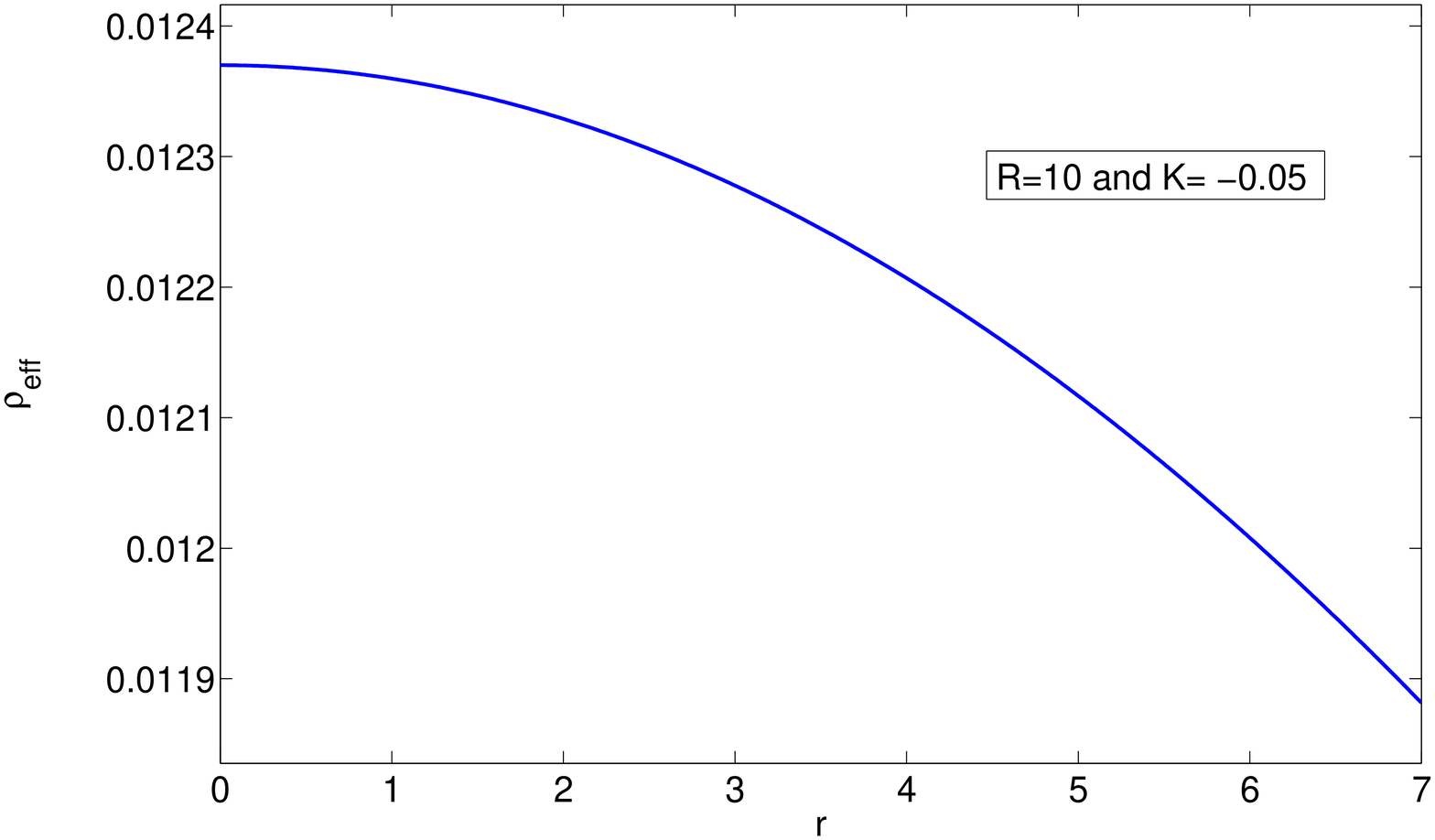}
       \caption{Effective density has been plotted against $r$ }
    \label{fig:3}
\end{figure}

\begin{figure}[htbp]
    \centering
        \includegraphics[scale=.3]{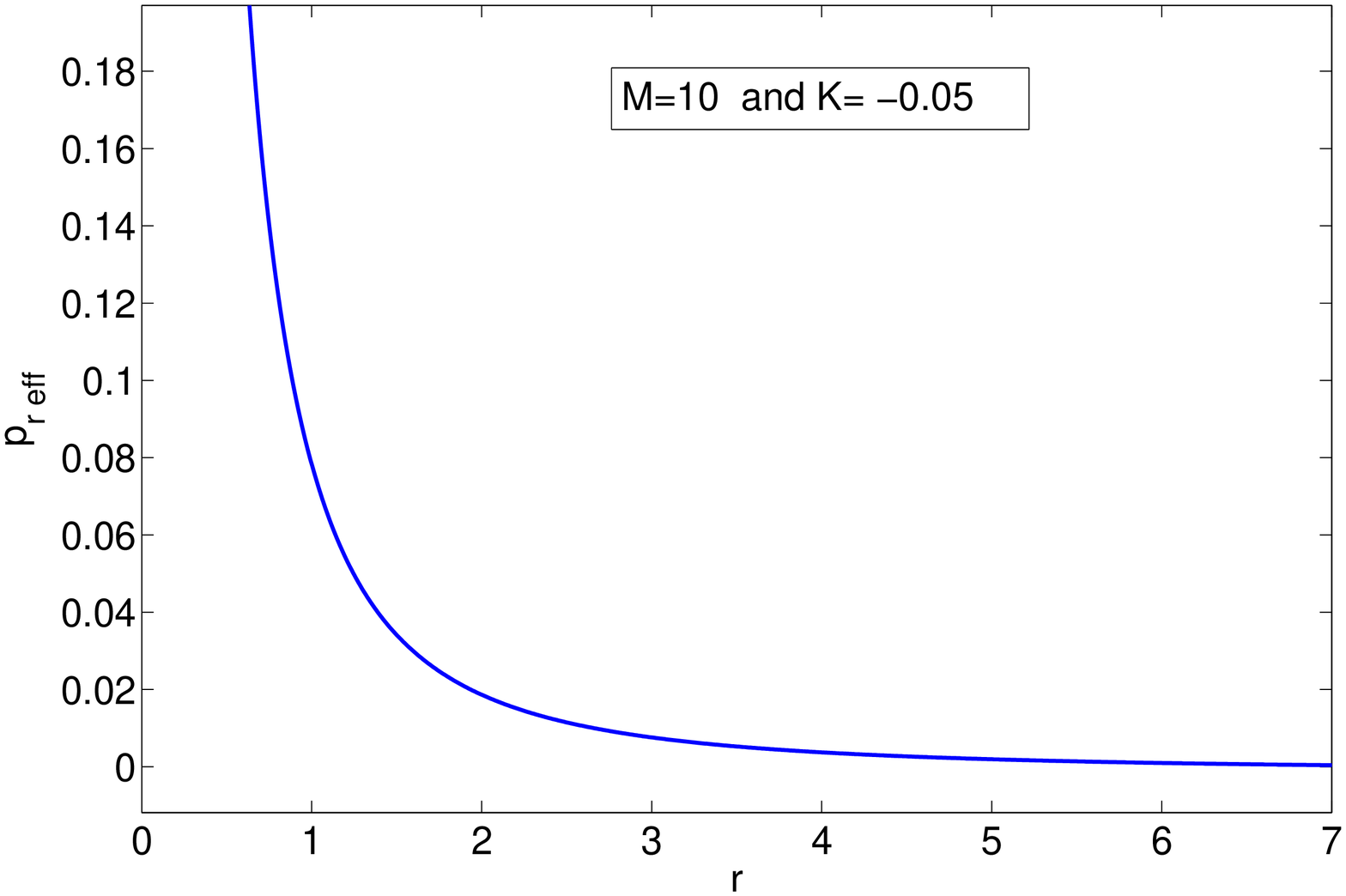}
       \caption{Effective radial pressure has been plotted against $r$ }
    \label{fig:3}
\end{figure}
The plot of effective density and effective radial pressure have been shown in $fig. 1$ and $fig. 2$ respectively. From these two figures we see that
both are monotonic decreasing function of 'r' and approaches to zero at the surface of the star.

\section{Physical Analysis}
The effective central density of the quintessence star is given by,
\begin{equation}
\rho_{0~eff}=\rho_{eff}(r=0)=\frac{3(1-K)}{8\pi R^{2}}
\end{equation}
\begin{equation}
\frac{d\rho_{eff}}{dr}=\frac{K(1-K)}{4\pi}\frac{r(5R^{2}-Rr^{2})}{(R^{2}-Kr^{2})^{3}}
\end{equation}
\begin{equation}
\frac{d^{2}\rho_{eff}}{dr^{2}}|_{r=0}=\frac{5K(1-K)}{4\pi R^{4}}
\end{equation}
So we see that effective density is regular at the center.The plot of $\frac{d\rho_{eff}}{dr}~ vs.~r$ has been shown in $fig.~3$. Both the expressions given in equation (35) and (36) is negative since we have chosen $K<0$ for our model.Which tells us that the effective density has maximum value at the center of the star.
\begin{figure}[htbp]
    \centering
        \includegraphics[scale=.33]{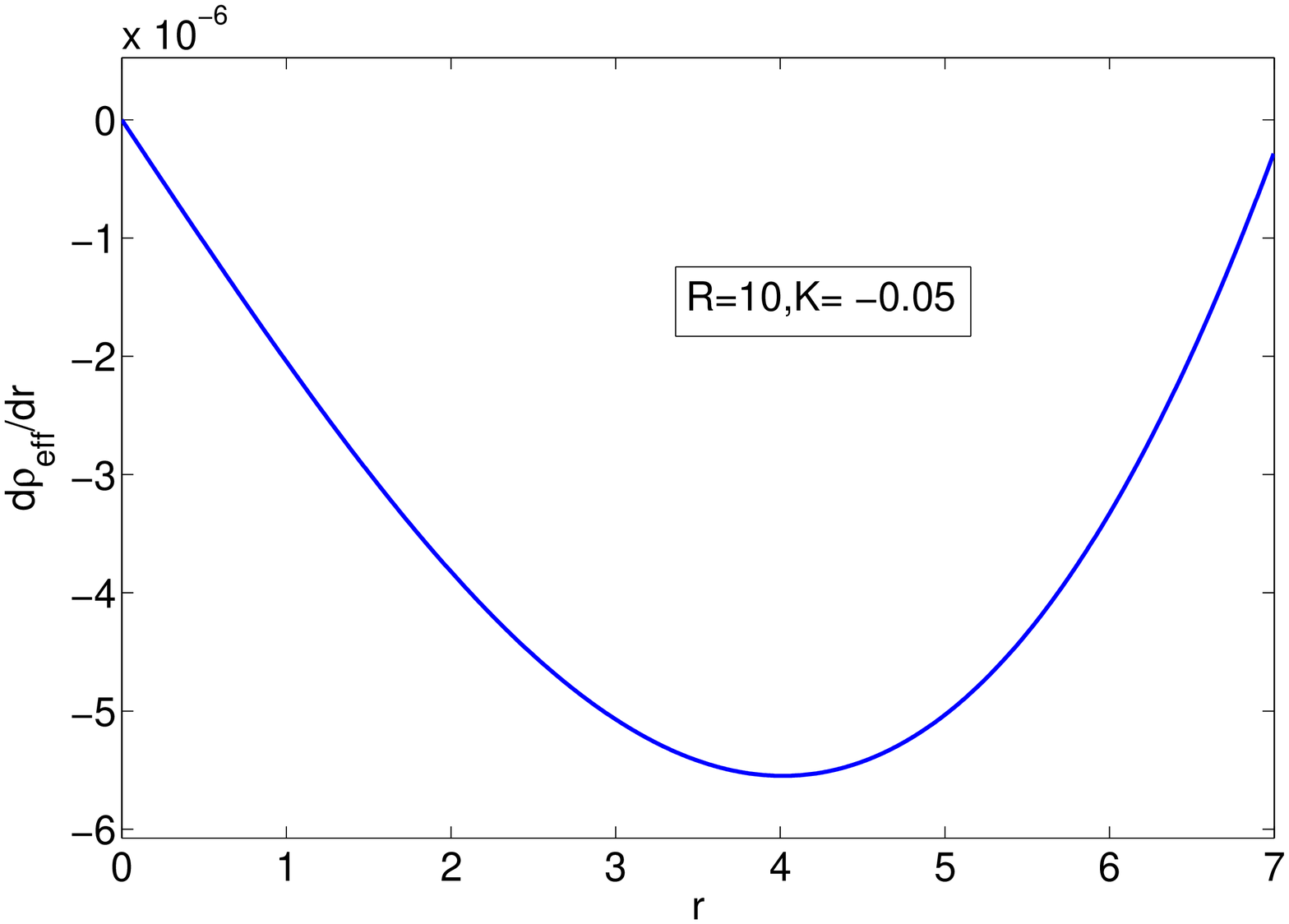}
       \caption{$\frac{d\rho_{eff}}{dr}$ has been plotted against $r$ }
    \label{fig:3}
\end{figure}

From the expression of $p_{r~eff}$ given in equation (32) we see that effective radial pressure is not regular at the center.
However
\begin{equation}
\frac{dp_{r~eff}}{dr}= -\frac{2\left[2R^{4}+K(3-K)r^{4}-4Kr^{2}R^{2}\right]}{r^{3}(R^{2}-Kr^{2})^{2}}~<0~~~
\end{equation}
The profile of $\frac{dp_{r~eff}}{dr}~vs.~r$ has shown in $fig.~4$.This figure once again verifies that $\frac{dp_{r~eff}}{dr}<0$ \\

\begin{figure}[htbp]
    \centering
        \includegraphics[scale=.3]{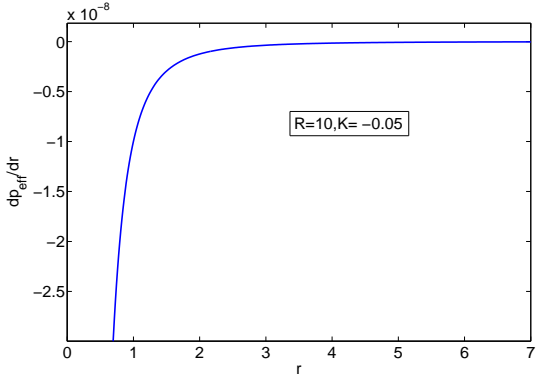}
       \caption{$\frac{dp_{eff}}{dr}$ has been plotted against $r$ }
    \label{fig:3}
\end{figure}

~~~~~The Anisotropic factor of our model of quintessence star is defined by
\[\Delta=(p_{t~eff}-p_{r~eff})\]
\begin{equation}
~~~=\frac{1}{8\pi}\left[\frac{1}{r^{2}}\frac{-R^{2}+(2-K)r^{2}}{R^{2}-Kr^{2}}-\frac{2(1-K)R^{2}}{(R^{2}-Kr^{2})^{2}}\right]
\end{equation}
and $\frac{2\Delta}{r}$ is termed as anisotropic force.

\begin{figure}[htbp]
    \centering
        \includegraphics[scale=.3]{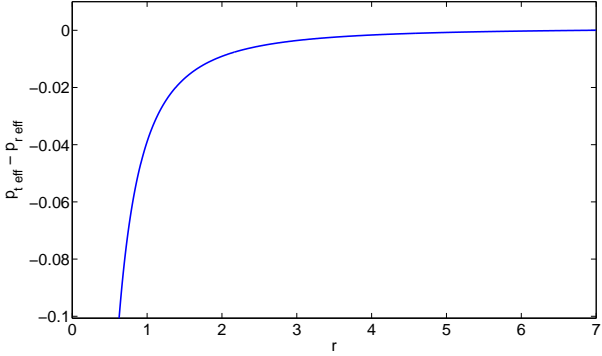}
       \caption{Anisotropic factor has been plotted against $r$ using 'R=10' and 'K= -0.05'}
    \label{fig:3}
\end{figure}
The profile of anisotropic factor has been given in $fig.~5$,which shows that $\Delta <0$ which implies that $p_t<p_r$.From here we can conclude that the force is attractive in nature.One can note that generally the quintessence field is repulsive in nature but for our model where we have considered the combine effect of ordinary matter and quintessence field the effective anisotropic force turns out as attractive.

\section{Exterior Spacetime And Matching Condition}
In this section we will match our interior solution of the quintessence star to the schwarzschild exterior solution  at the boundary $r=a$ outside the event horizon i.e $a>2M$\\

~where the exterior spacetime is described by the metric
\[ds^{2}=-\left(1-\frac{2M}{r}\right)dt^{2}+\left(1-\frac{2M}{r}\right)^{-1}dr^{2}\]
\begin{equation}
~~~~~~~~~~~~~~~~~~~~~~~~~~~+r^{2}(d\theta^{2}+\sin^{2}\theta d\phi^{2})
\end{equation}
Using the matching condition at the boundary we have
\begin{equation}
1-\frac{2M}{a}=C_2^{2}a^{2}
\end{equation}
and
\begin{equation}
\left(1-\frac{2M}{a}\right)^{-1}=\frac{R^{2}-Ka^{2}}{R^{2}-a^{2}}
\end{equation}
Solving  the above two equations we get,
\begin{equation}
C_2^{2}=\frac{1}{a^{2}}\left(1-\frac{2M}{a}\right)
\end{equation}
\begin{equation}
K=\frac{1-\left(\frac{2M}{a}\right)\left(\frac{R^{2}}{a^{2}}\right)}{1-\frac{2M}{a}}
\end{equation}

\section{TOV Equation}
To describe the static equilibrium let us consider the generalized Tolman-Oppenheimer-Volkov (TOV) equation which is represented by the formula
\begin{equation}
-\frac{M_G(\rho+p_r)}{r^{2}}e^{\frac{\lambda-\nu}{2}}-\frac{dp_r}{dr}+\frac{2}{r}(p_t-p_r)=0
\end{equation}
Where $M_G=M_G(r)$  is termed as is the effective gravitational mass inside the fluid sphere of radius 'r'and is defined by
\begin{equation}
M_G(r)=\frac{1}{2}r^{2}e^{\frac{\nu-\lambda}{2}}\nu'
\end{equation}
The above expression of $M_G(r)$ can be derived from Tolman-Whittaker mass formula\\
\begin{figure}[htbp]
    \centering
        \includegraphics[scale=.28]{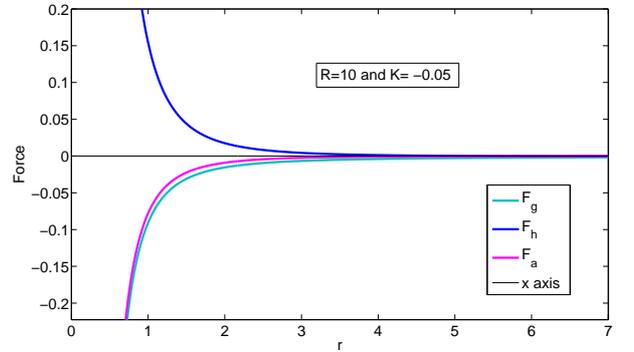}
       \caption{The system is in static equilibrium under three forces }
    \label{fig:3}
\end{figure}
Using the expression of equation $(45)$ in $(44)$ we obtain the modified TOV equation as,
\begin{equation}
F_g+F_h+F_a=0
\end{equation}
where
\begin{equation}
F_g=-\frac{\nu'}{2}(\rho_{eff}+p_{r~eff})
\end{equation}
\begin{equation}
F_h=-\frac{dp_r}{dr}
\end{equation}
\begin{equation}
F_a=\frac{2}{r}(p_{t~eff}-p_{r~eff})
\end{equation}
Where $F_g,F_h$ and $F_a$ are termed as gravitational, hydrostatics and anisotropic forces respectively of the system.From $Fig.~6$ we see that The system is in equilibrium under the above three forces.

\section{Energy Condition}
The null energy condition(NEC),weak energy condition (WEC),strong energy condition(SEC) is satisfied for our model if the following inequalities holds in the interior of the fluid sphere.
\begin{equation}
\rho_{eff}\geq0
\end{equation}
\begin{equation}
\rho_{eff}+p_{r~eff}\geq 0
\end{equation}
\begin{equation}
\rho_{eff}+p_{t~eff}\geq 0
\end{equation}
\begin{equation}
\rho_{eff}+p_{r~eff}+2p_{t~eff}\geq0
\end{equation}

\begin{figure}[htbp]
    \centering
        \includegraphics[scale=.3]{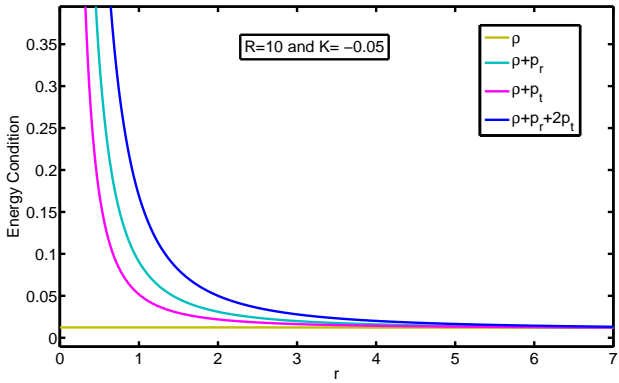}
       \caption{Energy Conditions have been plotted against $r$ }
    \label{fig:3}
\end{figure}

We will prove these inequalities with the help of graphical representation by choosing the arbitrary values to the parameters which has been given in $fig.~7$.From the figure we see that WEC,NEC and SEC are satisfied by our model.Since SEC is satisfied by our model so we can conclude that our spacetime does not contain any black hole.

\section{Stability}
For a physically acceptable model one must have the velocity of sound should be in the range $0<v^{2}=\frac{dp}{d\rho}\leq 1$\\

In case of anisotropy the radial $(v_{sr}^{2})$ and transverse $(v_{st}^{2})$ sound velocity can be obtained as
\begin{equation}
v_{sr}^{2}=\frac{dp_r}{d\rho}=m=0.4<1
\end{equation}
\begin{equation}
v_{st}^{2}=\frac{dp_t}{d\rho}
\end{equation}

\begin{figure}[htbp]
    \centering
        \includegraphics[scale=.35]{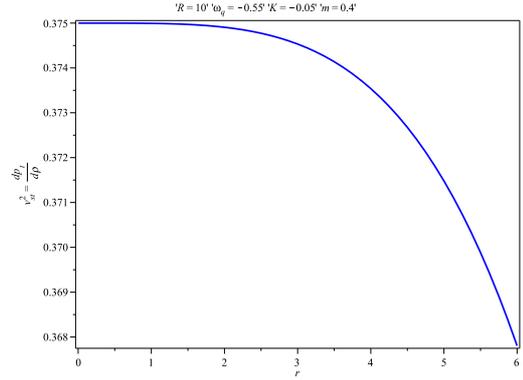}
       \caption{Transverse velocity has been shown against $r$ }
    \label{fig:3}
\end{figure}

From equation(54)we have $v_{sr}^{2}<1$ and $fig.~8$ shows that $0<v_{st}^{2}<1$ everywhere within the stellar configuration.According to Herrera's \cite{herrera} Cracking (or overturning) theorem for a potentially stable region one must have $v_{st}^{2}-v_{sr}^{2}<0$.From $fig.~9$ it is clear that our model satisfies this condition.So we conclude that our model is potentially stable.Moreover $0<v_{sr}^{2}\leq 1$ and $0<v_{st}^{2}<1$ therefore according to Andr\'{e}asson \cite{andreasson},$\left|v_{st}^{2}-v_{sr}^{2}\right|\leq 1$ which is also clear from $fig.~10$

\begin{figure}[htbp]
    \centering
        \includegraphics[scale=.38]{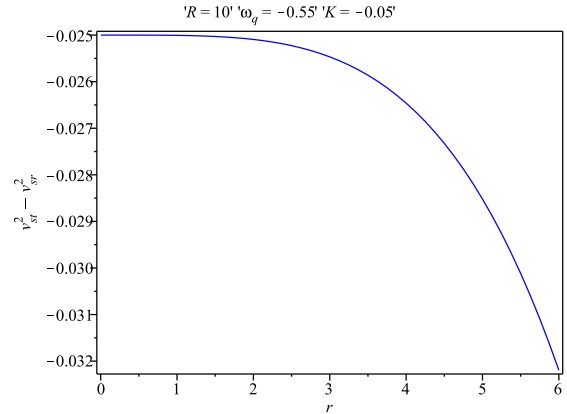}
       \caption{$v_{st}^{2}-v_{sr}^{2}$ has been plotted against $r$ }
    \label{fig:3}
\end{figure}

\begin{figure}[htbp]
    \centering
        \includegraphics[scale=.32]{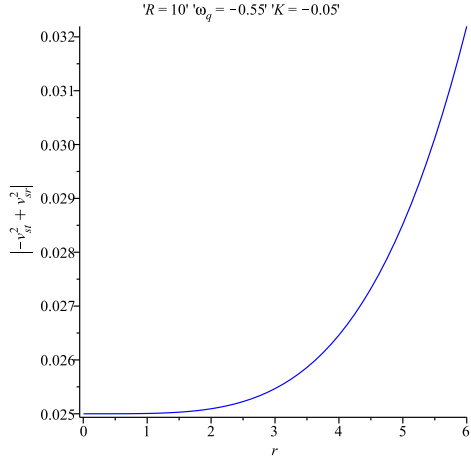}
       \caption{$\left|-v_{st}^{2}+v_{sr}^{2}\right|$ has been plotted against $r$ }
    \label{fig:3}
\end{figure}

\section{Some Features}

\subsection{Mass Radius Relation}
The mass function within the radius 'r'can be obtained as,
\begin{equation}
M_{eff}(r)=\int_0^{r}4\pi r^{2} \rho_{eff} dr=\frac{1-K}{2}\frac{r^{3}}{R^{2}-Kr^{2}}
\end{equation}
The profile of effective mass function has been given in fig.~11.For $r\rightarrow 0$,$m(r)\rightarrow 0$ which implies that mass function is regular at the center.

\begin{figure}[htbp]
    \centering
        \includegraphics[scale=.3]{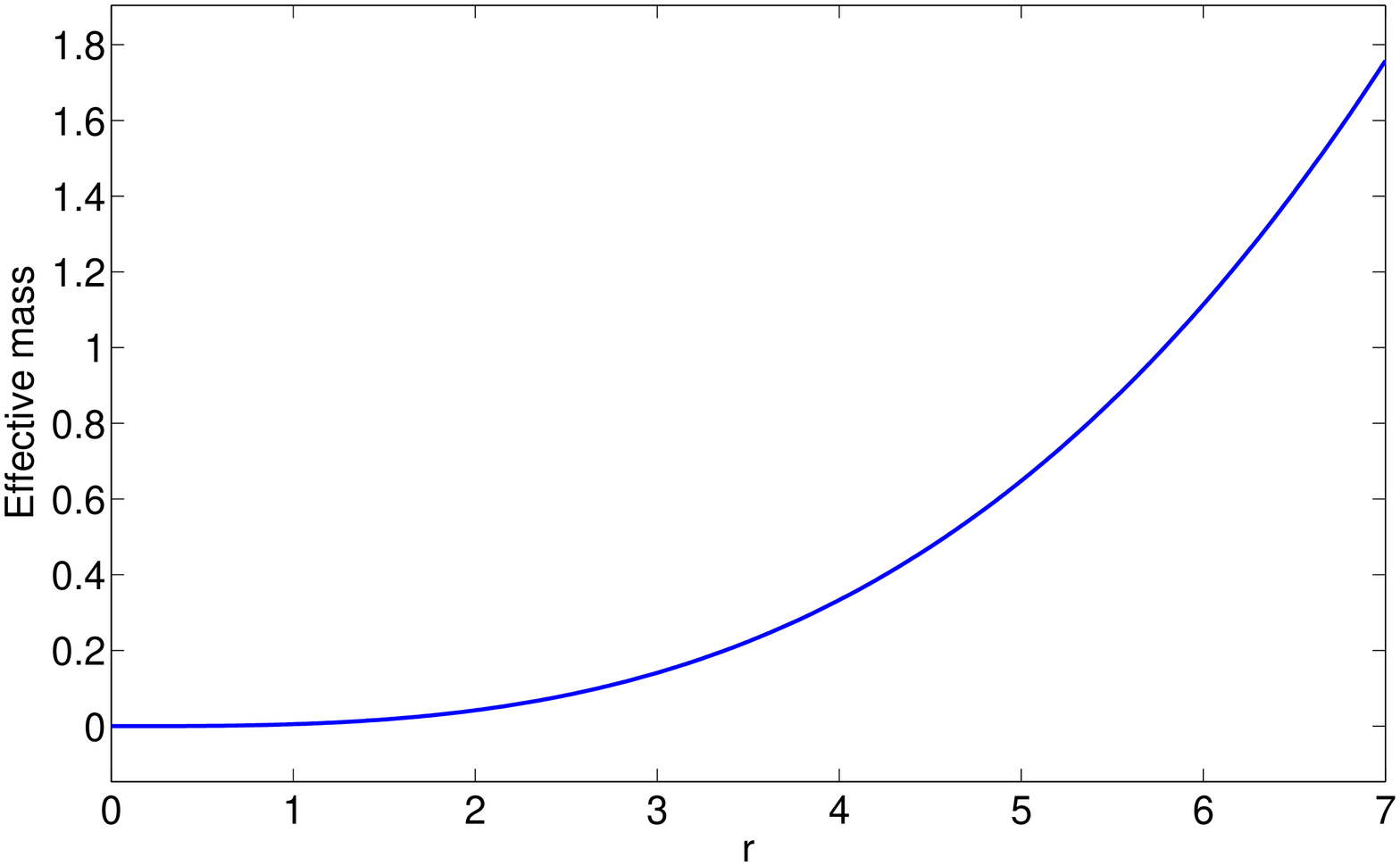}
       \caption{Effective mass function is plotted against $r$ }
    \label{fig:3}
\end{figure}
\subsection{Compactness}
The effective compactness of the star ${u_{eff}}(r)$ can be defined by,
\begin{equation}
{u_{eff}}=\frac{M_{eff}(r)}{r}=\frac{1-K}{2}\frac{r^{2}}{R^{2}-Kr^{2}}
\end{equation}
The profile of effective compactness of the star has been depicted in $fig.~12$
\begin{figure}[htbp]
    \centering
        \includegraphics[scale=.3]{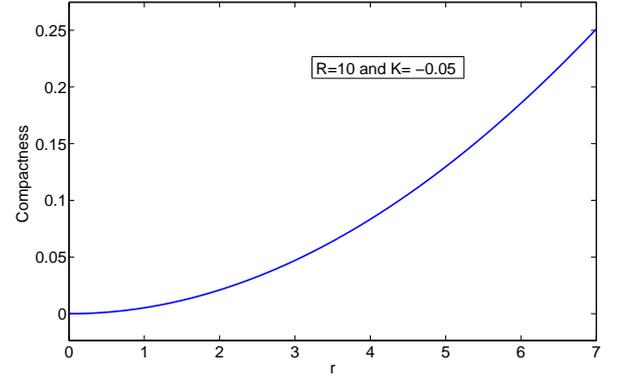}
       \caption{Effective Compactness has been plotted against $r$ }
    \label{fig:3}
\end{figure}

\subsection{Surface Redshift}
The redshift function ${Z}_{s~eff} $ can be defined by,
\begin{equation}
{Z}_{s~eff}=(1-2u_{eff})^{-\frac{1}{2}}-1=\left(\frac{R^{2}-r^{2}}{R^{2}-Kr^{2}}\right)^{-\frac{1}{2}}-1
\end{equation}

\begin{figure}[htbp]
    \centering
        \includegraphics[scale=.3]{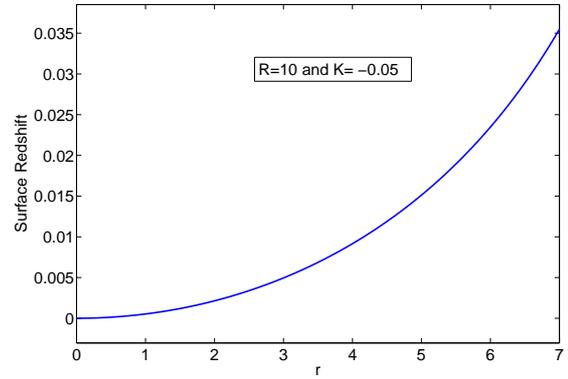}
       \caption{Surface redshift function has been plotted against $r$ }
    \label{fig:3}
\end{figure}
The profile of the effective redshift function is given in $fig.~13$

\section{Junction Condition}
In section $6.$ we have matched our interior spacetime to the exterior Schwarzschild at the boundary $r=a$.It is obvious that the metric coefficients are continuous at $r=a$,but it does not ensure that their derivatives are also continuous at the junction surface.To take care of this let us consider the Darmois-Israel\cite{israel1,israel2} formation to determine the surface stresses at the junction boundary.The intrinsic surface stress energy tensor $S_{ij}$ is given by Lancozs equations in the following form
\begin{equation}
S^{i}_{j}=-\frac{1}{8\pi}(\kappa^{i}_j-\delta^{i}_j\kappa^{k}_k)
\end{equation}
The second fundamental form is given by,
\begin{equation}
K_{ij}^{\pm}=-n_{\nu}^{\pm}\left[\frac{\partial^{2}X_{\nu}}{\partial \xi^{1}\partial\xi^{j}}+\Gamma_{\alpha\beta}^{\nu}\frac{\partial X^{\alpha}}{\partial \xi^{i}}\frac{\partial X^{\beta}}{\partial \xi^{j}} \right]|_S
\end{equation}
and the discontinuity in the second fundamental form is given by,
\begin{equation}
K_{ij}=K_{ij}^{+}-K_{ij}^{-}
\end{equation}
where $n_{\nu}^{\pm}$ are the unit normal vector defined by,
\begin{equation}
n_{\nu}^{\pm}=\pm\left|g^{\alpha\beta}\frac{\partial f}{\partial X^{\alpha}}\frac{\partial f}{\partial X^{\beta}}  \right|^{-\frac{1}{2}}\frac{\partial f}{\partial X^{\nu}}
\end{equation}
with $n^{\nu}n_{\nu}=1$.Where $\xi^{i}$ is the intrinsic coordinate on the shell.$+$ and $-$ corresponds to exterior i.e, Schwarzschild  spacetime and interior (our) spacetime respectively.\\
The non-trivial components of the extrinsic curvature are given by
\begin{equation}
K_{\tau}^{\tau~+}=\frac{\frac{M}{a^{2}}+\ddot{a}}{\sqrt{1-\frac{2M}{a}+\dot{a}^{2}}}
\end{equation}
\begin{equation}
K_{\tau}^{\tau~-}=\frac{\frac{-a(1-K)R^{2}}{(R^{2}-Ka^{2})^{2}}+\ddot{a}}{\sqrt{\frac{R^{2}-a^{2}}{R^{2}-Ka^{2}}+\dot{a}^{2}}}
\end{equation}
and
\begin{equation}
K_{\theta}^{\theta~+}=\frac{1}{a}\sqrt{1-\frac{2M}{a}+\dot{a}^{2}}
\end{equation}
\begin{equation}
K_{\theta}^{\theta~-}=\frac{1}{a}\sqrt{\frac{R^{2}-a^{2}}{R^{2}-Ka^{2}}+\dot{a}^{2}}
\end{equation}

Considering the spherical symmetry of the spacetime surface stress energy tensor can be written as $S^{i}_j=diag(-\sigma,\mathcal{P})$.Where $\sigma$ and $\mathcal{P}$ is the surface energy density and surface pressure respectively.

\[\sigma=-\frac{1}{4\pi a}\left[\sqrt{e^{-\lambda}} \right]_{-}^{+}\]
\begin{equation}
~~~=-\frac{1}{4\pi a}\left[\sqrt{1-\frac{2M}{a}+\dot{a}^{2}}-\sqrt{\frac{R^{2}-a^{2}}{R^{2}-Ka^{2}}+\dot{a}^{2}}~\right]
\end{equation}

\[\mathcal{P}=\frac{1}{8\pi a}\left[\left\{1+\frac{a\nu'}{2}\right\}\sqrt{e^{-\lambda}}\right]_{-}^{+}\]
\begin{equation}
=\frac{1}{8\pi a}\left[\frac{1-\frac{M}{a}}{\sqrt{1-\frac{2M}{a}}}-2\sqrt{\frac{R^{2}-a^{2}}{R^{2}-Ka^{2}}}\right]
\end{equation}

Hence we can match our interior spacetime to the exterior Schwarzschild spacetime in presence of a thin shell.

The mass of the thin shell is given by
\begin{equation}
m_s=4\pi a^{2}\sigma
\end{equation}
From $(67)$ and $(69)$ one can obtain
\begin{equation}
M=\frac{a^{3}}{2}\left[\frac{1-K}{R^{2}-Ka^{2}}-m_s^{2}\right]+2am_s\sqrt{\frac{R^{2}-a^{2}}{R^{2}-Ka^{2}}}
\end{equation}
Which gives the mass of the quintessence star in terms of the thin shell mass.\\

~~~~Next we will discuss about the evolution identity given by $[T_{\mu\nu }n^{\mu}n^{\nu}]_{-}^{+}=\overline{K}^{i}_{j}S^{j}_{i}$
where
\[\overline{K}^{i}_{j}=\frac{1}{2}\left(K_{j}^{i+}+K_{j}^{i-}\right)\]
 which gives,
\[p_r+\frac{(\rho+p_r)\dot{a}^{2}}{\frac{R^{2}-a^{2}}{R^{2}-Ka^{2}}}\]
\[=-\frac{1}{2a}\left(\sqrt{1-\frac{2M}{a}+\dot{a}^{2}}+\sqrt{\frac{R^{2}-a^{2}}{R^{2}-Ka^{2}}+\dot{a}^{2}}\right)\mathcal{P}\]
\begin{equation}
+\frac{1}{2}\left(\frac{\frac{M}{a^{2}}+\ddot{a}}{\sqrt{1-\frac{2M}{a}+\dot{a}^{2}}}+\frac{\frac{-a(1-K)R^{2}}{(R^{2}-Ka^{2})^{2}}
+\ddot{a}}{\sqrt{\frac{R^{2}-a^{2}}{R^{2}-Ka^{2}}+\dot{a}^{2}}}\right)\sigma
\end{equation}
For a static solution $a_0$ from equation $(71)$(assuming $\dot{a}=0=\ddot{a}$)~one can obtain
\[p_r(a_0)=-\frac{1}{2a_0}\left(\sqrt{1-\frac{2M}{a_0}}+\sqrt{\frac{R^{2}-a_0^{2}}{R^{2}-Ka_0^{2}}}~\right)\mathcal{P}\]
\begin{equation}
+\frac{1}{2}\left(\frac{\frac{M}{a_0^{2}}}{\sqrt{1-\frac{2M}{a_0}}}+\frac{\frac{-a_0(1-K)R^{2}}{(R^{2}-Ka_0^{2})^{2}}}
{\sqrt{\frac{R^{2}-a_0^{2}}{R^{2}-Ka_0^{2}}}}\right)\sigma
\end{equation}
The above equation relates the radial pressure $(p_r)$ of the quintessence star with the surface pressure $(\mathcal{P})$ and surface density $(\sigma)$ of the thin shell.

\section{Discussion and concluding remarks}
In the present paper we have proposed a new model of superdense star by choosing Vaidya-Titekar spacetime which admits CKV in presence of quintessence field which is characterized by a parameter $\omega_q$ with $-1<\omega_q<-\frac{1}{3}$. For our model the effective density is regular at the center but the radial and transverse pressure suffers from central singularity like other CKV model.The profile of both the density function and radial pressure are monotonic decreasing which indicates that the density and radial pressure of the star is maximum at the center and it decreases from the center to the surface of the star.The mass function is regular at the center and the maximum allowable ratio of mass to radius is $0.25<\frac{4}{9}$ which lies in the Buchdahl \cite{buchdahl} limit.For our model the radial and transverse speed of sound is less than $1$,which gives the stability condition.According to Herrera \cite{herrera} concept if for a model radial speed of sound is greater than the transverse speed of sound ,the model is potentially stable.With the help of graphical representation we have shown that $v_{sr}^{2}-v_{st}^{2}>0$.So our model is potentially stable.All the energy conditions is satisfied inside the fluid sphere.Our model is also in static equilibrium under anisotropic,gravitational and hydrostatic forces.We have matched our interior solution to the exterior
Schwarzschild metric in the presence of thin shell and also obtained the mass of the quintessence star in terms of the thin shell mass.A relation among the radial pressure ,surface pressure and surface density has been obtained.

\end{document}